\documentclass[twocolumn,showpacs,preprintnumbers,amsmath,amssymb,eqsecnum]{revtex4}
\usepackage{graphicx}
\begin{document}
\preprint{IMAFF-RCA-010-01-02}
\title{Multidimensional quantum cosmic models: New solutions and gravitational waves}

\author{Pedro F. Gonz\'{a}lez-D\'{\i}az and Alberto Rozas-Fern\'{a}ndez}
\affiliation{Colina de los Chopos, Centro de F\'{\i}sica ``Miguel
A.
Catal\'{a}n'', Instituto de F\'{\i}sica Fundamental,\\
Consejo Superior de Investigaciones Cient\'{\i}ficas, Serrano 121,
28006 Madrid (SPAIN).}
\date{\today}
\begin{abstract}
This paper contains a discussion on the quantum cosmic models,
starting with the interpretation that all of the accelerating
effects in the current universe are originated from the existence
of a nonzero entropy of entanglement. In such a realm, we obtain
new cosmic solutions for any arbitrary number of spatial
dimensions, studying the stability of these solutions, so as the
emergence of gravitational waves in the realm of the most general
models.
\end{abstract}


\maketitle

\section{Introduction}

No doubt, one of the greatest problems that the twenty-one century
has inherited from the last decade of the twenty century and
remains up today still unsolved is the so-called problem of the
dark energy by which it appears to be completely impossible to
account for the observed current accelerated expansion of the
universe (which has been confirmed by several kinds of
observational checking) by only using classical general relativity
and its conventional cosmological solutions [1]. So far, two
streams have been mostly followed to try to solve that problem by
implementing the necessary cosmic repulsive force either by
including a scalar field, called quintessence [2] or k-essence
[3], or modifying the Hilbert-Einstein gravity action by adding
suitable extra terms to it [4]. None of them appears to be
consistent enough or could fulfill all observational requirements
[5].

A theory which is self-consistent and agrees with all
observational data has been recently proposed [6-9]. It is based
on the assumption that all the accelerating effects come from the
very quantum-entangled nature of the current universe [9,10]. In
such a framework one can get essentially two relevant quantum
solutions both of which can be seen as quantum perturbations to
the de Sitter space [8,9], which is recovered in the classical
limit where $\hbar\rightarrow 0$. It has been also shown that out
from these two possible solutions only one of them satisfies the
second law of thermodynamics [9], and hence is physically
meaningful. It corresponds to a phantom universe [11] in that the
parameter of the equation of state gets always on values which are
less than -1, but does not show any violent instability [12] nor
the sort of inconsistency coming from having a negative kinetic
term for the scalar field - in fact, these models do not actually
contain any scalar or other kinds of vacuum fields in their final
equations. It is for these reasons that such a cosmic model has
been also denoted as [8,9] {\it benigner} phantom model. On the
other hand, given that de Sitter space is stable to scalar
perturbations and that vectorial perturbations are in any case
pure gauge [13], since the considered solutions can be regarded as
nothing but scalar perturbations on de Sitter space, we ought to
conclude that they are stable under such scalar harmonically
symmetric perturbations and that those with vectorial character
are also pure gauge in the case of the quantum solutions.

However, one cannot still be sure that the solution which has been
chosen as the most physically relevant is stable under
semiclassical and tensorial perturbations leading to gravitational
waves. In this paper we shall study these two kinds of
perturbations, showing that they in fact follows the same
stability pattern as that of the de Sitter space. Our developments
are made on a generalization of the quantum closed models to any
number of dimensions and to the case in which a black hole is
inserted in the space-time. Throughout this paper we will use
natural unit so that $c=\hbar=1$, unless otherwise stated.

The paper can be outlined as follows. In Sec. II we very briefly
review the quantum cosmic models, its origin and interpretation.
The generalization of such models both to higher dimensions and to
new models which contain a black hole are considered in Sec. III,
together with a study of the static counterparts of such
generalizations. Tensorial and semiclassical perturbations on the
resulting closed cosmic models are studied in Sec. IV, briefly
concluding in Sec. V.

\section{quantum cosmic models}
As it was already advanced in the Introduction, the quantum cosmic
models provide us with a dynamical cosmological scenario for the
current evolution of the universe which uses just general
relativity with or without a cosmological constant and the
sharpest aspects of quantum mechanics, without inserting any kind
of vacuum fields or introducing any extra terms in the
Hilbert-Einstein gravitational action [6-9]. Such aspects can
alternatively be viewed either as a sub-quantum potential or as
due to the existence of an entanglement entropy for the currently
accelerating universe. Because in Refs. [8] the first of these two
equivalent interpretations was reviewed, we here summarize the
main points of the quantum cosmic models by using also the second
one. Actually, since the holographic version of the quantum cosmic
models comes quite naturally from such models in terms of the
physically consistent interpretation that the holographic screen
coincides with the Hubble horizon [8], and at least one of the
models (precisely that which satisfies the second law of
thermodynamics) is formally equivalent with the Barrow's hyper
inflationary model [14], we shall now briefly comment on this
interpretation.

Let us interpret from the onset the quantity $E_{Ent.}=a^3 V_{SQ}$
as the total entanglement energy [10] of a universe with scale
factor $a\equiv a(t)$ and whose matter-radiation content can be
characterized by a sub-quantum potential density $V_{SQ}$ [6-9].
In such a case, the latter potential can simply be taken to
describe the entanglement energy density of the universe, which we
will denote as $\epsilon_{Ent.}\equiv E_{Ent.}/a^3$. In this way,
because of the additiveness of the entanglement entropy, one can
add up [15] the contributions from all existing individual fields
in the observable universe in such a way that the entropy of
entanglement $S_{Ent.}=\beta R_H^2$, with $\beta$ a constant that
includes an account for the spin degrees of freedom of the quantum
fields in the Hubble observable volume of radius $R_H$ and a
numerical constant of order unity [15]. On the other hand, the
presence of a boundary at the horizon leads us to conclude that
the entanglement energy ought to be proportional to the radius of
the associated spherical volume, that is $E_{€nt.}=\alpha R_H$,
with $\alpha$ a given constant, and hence again $S_{Ent.}=\beta
R_H^2$.

It is worthy to notice that we can then take the temperature
derived from the thermodynamics of the quantum cosmic model
respecting the second law with $a=a_+$ as the entanglement
temperature so that $T(a_+)=E_{Ent.}/k_B$. Using now the general
expression [10]
\[dE_{Ent.}=T_{Ent.}dS_{Ent.} ,\]
where $T_{Ent.}=(2\pi R_H)^{-1}$ is the Hawking-Gibbons
temperature, we consistently recover once again the expression
$S_{Ent.}=\beta R_H^2$ for $\alpha=\beta/\pi$. That result is also
consistent with the most natural holographic expression [8]
$\rho=3/(8\pi GR_H^2)$, described in terms of the Hubble horizon
in the quantum cosmic models.

The models based on a sub-quantum potential and those derived from
the existence of an entangled energy in the universe are all
originated from the consideration of a Langrangian density given
by [6-8]
\[L=V(\phi)\left[E(x,\kappa)-\sqrt{1-\dot{\phi}^2}\right] ,\]
where $\phi$ is an auxiliary scalar field,
$V(\phi)=\tilde{V(\phi)}/a^3$ is its associated potential energy
density, $E(x,\kappa)$ is the elliptic integral of the second
kind, with $x=\arcsin\sqrt{1-\dot{\phi}^2}$ and
$\kappa=\sqrt{1-V_{SQ}^2/v(\phi)}\equiv\sqrt{1-\epsilon_{Ent.}^2/V(\phi)}$,
where $V_{SQ}=\tilde{V}_{SQ}/a^3$ is the sub-quantum potential
density.

\begin{figure}
\includegraphics[width=.9\columnwidth]{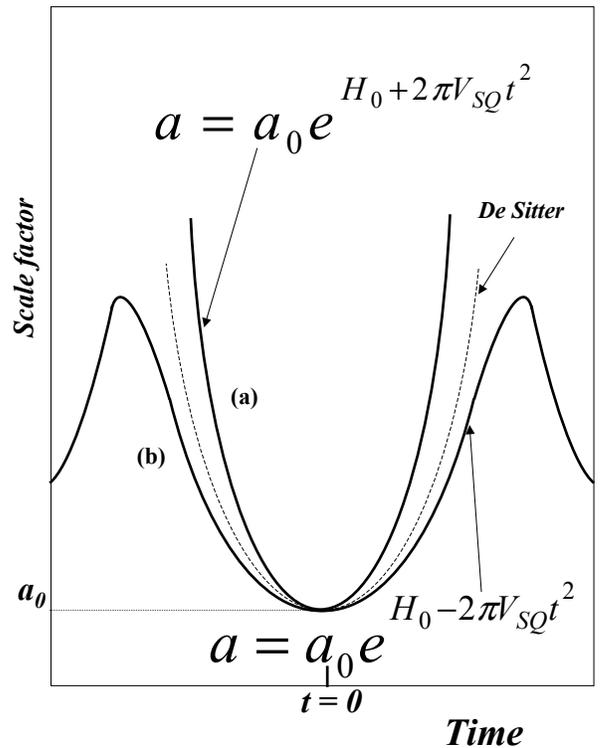}
\caption{\label{fig:epsart} Cosmic solutions that result from the
introduction of an energy density of entanglement
$\epsilon_{Ent.}\equiv V_{SQ}$, where $V_{SQ}$ is the sub-quantum
potential. Solution (a) goes like in De Sitter space with the same
$H_0$, but with higher acceleration. Solution (b) corresponds to
the case where $H_0^2>4\pi \epsilon_{Ent.}$ and represents a
universe which is initially expanding in an accelerated way (at a
rate slower than in De Sitter space with the same $H_0$), then
expands in a decelerated way for a while to finally contract
toward a zero radius as $t\rightarrow\infty$. On the figure we
have used units such that $\hbar=c=G=1$}
\end{figure}

The above Lagrangian density vanishes in the classical limit
$\hbar\rightarrow 0$. Adding physically reasonable regularity
conditions for $\ddot{\phi}$ [7,8] we get $\dot{\phi}^2 =1$. Thus,
the use of the above Lagrangian density and the Lagrange
equations, together with the final formulae derived in Ref. [8]
then yield the following general expressions for the energy
density and pressure
\begin{equation}
\rho=\frac{p}{w(t)}= 6\pi
G\left(\dot{H}^{-1}HV_{SQ}\right)^2\equiv 6\pi
G\left(\dot{H}^{-1}H\epsilon_{Ent.}\right)^2 ,
\end{equation}
and for the time -dependent parameter of the equation of state
$p=w(t)\rho$
\begin{equation}
w(t)=-\left(1+\frac{2\dot{H}}{3H^2}\right),
\end{equation}
with
\begin{equation}
\dot{H}=\pm 4\pi GV_{SQ}\equiv \pm 4\pi G\epsilon_{Ent.}
\end{equation}
\begin{equation}
H=\pm 4\pi G V_{SQ} t+H_0\equiv \pm 4\pi G \epsilon_{Ent.}t+H_0 ,
\end{equation}
and the set of cosmic solutions
\begin{equation}
a_{\pm}=a_0\exp \left(Ht\pm 2\pi GV_{SQ} t^2\right)\equiv a_0\exp
\left(Ht\pm 2\pi G\epsilon_{Ent.}t^2\right).
\end{equation}
In Eqs. (2.2)-(2.5) $H=\dot{a}/a$, $H_0$ is an integration
constant playing the role of a cosmological term, $H_0=
\Lambda^{1/2}$, and $a$ is the scale factor with $a_0$ its minimum
value at time $t=0$. Eqs. (2.3) - (2.5) are valid only for
sufficiently large $t$ or large $H_0$. From the set of solutions
implied by Eq. (2.5) we shall disregard from the onset the one
corresponding to $H_0=0$ and $t=\sqrt{\frac{\ln(a_0/a)}{2\pi
GV_{SQ}}}$ as it would predict the unphysical case of a universe
which necessarily is currently contracting. The chosen solutions
are depicted in Fig.1 as compared to the usual flat de Sitter
solution.

Actually, any account for the interpretation based on the cosmic
entangled energy density $\epsilon_{Ent.}$ can straightforwardly
be obtained from the corresponding account given in Refs. [6-9]
given in terms of the sub-quantum potential, by simply replacing
the sub-quantum potential energy density $V_{SQ}$ for
$\epsilon_{Ent.}$ in all reasoning and formulae. Two points should
be stressed now nevertheless. On the one hand, it is worth
remarking that one would not expect $\tilde{V}_{SQ}$ to remain
constant along the universal expansion, but to steadily increase
like the volume of the universe $V=a^3$ does [8]. In order for
obtaining the above relevant solutions one then must realize that
it is the sub-quantum potential energy density
$V_{SQ}=\tilde{V}_{SQ}/a^3$ appearing in the above Lagrangian
density what should then be expected to remain constant at all
cosmic times and, since we have consistently taken [8] $a^3
V_{SQ}$ to be the total entanglement energy of the universe, the
right-hand-side of solutions (2.5) appear to be correct, too. On
the other hand, the scenarios we are considering are meant to
evade or at least ameliorate he cosmic coincidence problem in the
following sense. Whereas the sub-quantum potential density has to
be chosen small enough to produce a sufficient late time
domination, at the same time no problematic fine-tuning would be
required for $\tilde{V}_{SQ}$ which must take on varying nearly
arbitrary large values because of the very large values of the
scale factor during the cosmic acceleration. In addition, since in
the present scenarios all existing matter particles and fields
should be associated with a sub-quantum potential (or entanglement
entropy) and one can show [9] that that potential (or entangled
energy) would make the effective mass of particles and field to
vanish precisely at the coincidence time, then a cosmic system can
be most naturally allowed where the matter dominance phase is
followed by the accelerating expansion without any conceptual
problem.

Current data based on a variety of observations [5] appear to
point out to a present value for the parameter of the equation of
state $w=-1$, with a bias toward slightly smaller values, that is
to say, currently $w$ can possibly be less than -1 by a very small
amount. This is actually the case that corresponds to the so
called phantom energy [11], a form of dark energy which shows two
main fundamental problems, a negative kinetic term in the
Lagrangian and a fatal singularity in the finite future [11] which
is associated with violent instabilities [12] and classical
violation of the dominant energy condition. While solution $a_+$,
(a) in Fig. 1, would approach the observational data as closely as
we want, it does not show any of the problems which have been
ascribed to phantom energy. Moreover, the given solution appears
[8]: (i) to be stable, (ii) having suitable thermodynamic
properties that consistently generalize those of the de Sitter
space and (iii) entails a admissible residual quantum violation of
the dominant energy condition ($\rho+p=-\epsilon$) leading to
consistent quantum wormhole solutions; on the other hand, (iv)
that description admits a most natural holographic extension where
the holographic screen is placed at the Hubble horizon, and (v) it
entails a perturbed metric which is no longer static but
consistently reduces to static de Sitter metric in the limit where
$w\rightarrow -1$ [8].

It it worth noticing finally that what actually matters in the
models dealt with in this section is that some quantum-mechanical
effects, which originally sub-dominated in the matter-dominated
phase, eventually started driving cosmic acceleration.

\section{Generalized cosmic solutions}

It has been already seen that the quantum cosmic solutions can be
seen as either some generalizations from the flat version of de
Sitter space or, if $V_{SQ}$ is sufficiently small, such as it
appears to actually be the case, as perturbations of that de
Sitter space. Since most of such models correspond to equations of
state whose parameter is less than -1, such as it was mentioned
before, they are also known as {\it benigner} phantom cosmic
models [6-9]. In this section we shall derive even more general
expressions for these quantum cosmic solutions by (i) considering
the similar generalizations or perturbations of the hyperbolic
version of the de Sitter space, and (ii) using a $d$-dimensional
manifold. Actually, some observational data have implied that our
universe is not perfectly flat and recent works [17,18]
contemplate the possibility of the universe having spatial
curvature. Thus, although WMAP alone abhors open models, requiring
$\Omega_{total}\equiv\Omega_{m}+\Omega_{\Lambda}=
1-\Omega_{k}\geq0.9$ (95\%), closed model with $\Omega_{total}$ as
large as 1.4 are still marginally allowed provided that the Hubble
parameter $h\sim0.3$ and the age of the Universe $t_{0}\sim20
Gyr$. The combinations of the WMAP plus the SNIa data or the
Hubble constant data also imply the possibility of the closed
universe, giving curvature parameters $k=-0.011 \pm 0.012$ and $k
= -0.014 \pm 0.017$, respectively [17], although the estimated
values are still consistent with the flat FRW world model.
Moreover, in Ref. [19] it is said that the best fit closed
universe model has $\Omega_{m}=0.415$, $\Omega_{\Lambda}=0.630$
and $H_{0}=55 kms^{-1}Mpc^{-1}$ and is a better fit to the WMAP
data alone than the flat universe model $(\Delta\chi_{eff}^{2}=2$.
However, the combination of WMAP data with either SNe data,
large-scale structure data or measurements of $H_{0}$ favors
models with $\Omega_{K}$ close to 0.

The $d$-dimensional de Sitter space has already been considered
elsewhere [16]. Here we shall extend it to the also maximally
symmetric space whose spacetime curvature is still negative
(positive Ricci scalar) but no longer constant. Our spacetime will
be solution of the Einstein equation
\begin{equation}
R_{ab}=t_{ab} ,\;\; a,b=0,1,...(d-1) ,
\end{equation}
with
\begin{equation}
t_{ab}=\left(H\pm \hbar\xi t\right)^2 g_{ab} ,
\end{equation}
where $H^2=\Lambda/(d-1)$ is a cosmological constant and the
constant $\hbar\xi$ generalizes the sub-quantum potential
considered in the quantum cosmic models described in Sec. II. We
notice that in the classical limit $\hbar\rightarrow 0$ the above
definition becomes that of the usual $d$-dimensional de Sitter
space. We shall restrict ourselves in this paper to the case in
which our generalized $d$-dimensional de Sitter space can still be
visualized as a $d+1$ hyperboloid defined as [20]
\begin{equation}
-x_0^2 +\sum_{j=1}^d x_j^2 = H^{-2} .
\end{equation}
This $(d+1)$-dimensional hyperboloid is embedded in E$^{d+1}$, so
that the most general expression of the metric for our extended
quantum-corrected solutions is provided by the metric induced in
this embedding, that is
\begin{equation}
ds^2 = -dx_0^2 +\sum_{j=1}^d dx_j^2 ,
\end{equation}
which has the same topology and invariance group as the
$d$-dimensional de Sitter space [16].

This metric can now be exhibited in coordinates
$\Theta_{\pm}=t(1\pm\hbar\xi t/H)\epsilon (\mp
H_0/(4\hbar\xi),\pm\infty)$ (notice that our solutions then only
covers a portion of the de Sitter time, while $t \epsilon
(-\infty,+\infty)$), $\psi_{d-1}, \psi_{d-2}, ... \psi_2 \epsilon
(0,\pi)$, $\psi_1 \epsilon (0,2\pi)$, defined by
\[x_d
=H^{-1}\cosh(H\Theta)\sin\psi_{d-1}\sin\psi_{d-2}...\sin\psi_2\cos\psi_1\]

\[x_{d-1}
=H^{-1}\cosh(H\Theta)\sin\psi_{d-1}\sin\psi_{d-2}...\sin\psi_2\sin\psi_1\]
\[x_{d-2}
=H^{-1}\cosh(H\Theta)\sin\psi_{d-1}\sin\psi_{d-2}...\cos\psi_2\]
\begin{equation}
\end{equation}
\[x_1 =H^{-1}\cosh(H\Theta)\cos\psi_{d-1}\]
\[x_0=H^{-1}\sinh(H\Theta) , \]
which should be referred to as either time $\Theta_+$ or time
$\Theta_-$. In terms of these coordinates metric (3.4) splits into
\begin{eqnarray}
&&ds_{\pm}^2 = -\left(1\pm\frac{2\hbar\xi t}{H}\right)^2
dt^2\nonumber\\ &&+H^{-2}\cosh^2 \left[t\left(H\pm\hbar\xi
t\right)\right]d\Omega_{d-1}^2 ,
\end{eqnarray}
where $d\Omega_{d-1}^2$ is the metric on the $(d-1)$-sphere.
Metric (3.6) is a closed $(d-1)$-dimensional
Friedmann-Robertson-Walker metric whose spatial sections are
$(d-1)$-spheres of radius $H^{-1}\cosh(H\Theta)$. The coordinates
defined by Eqs. (3.5) describe two closed quantum cosmic spaces,
$B_{\pm}$, which interconvert into each other at $t=0$. $B_+$
first steadily contracts until $t=0$ where it converts into $B_-$
to first expand up to a finite local maximum value at
$t=H/(2\hbar\xi)$, then contract down to $a_0$ at
$t=H/(\hbar\xi)$, expanding thereafter to infinite. $B_-$ would
first contract until $t=-H/(\hbar\xi)$, then expand up to reach a
local maximum at $t=-H/(2\hbar\xi)$, to contract again until
$t=0$, where it converts into $a_+$ which will steadily expand
thereafter to infinite.

In terms of the conformal times $\eta_{\pm}=\int
d\Theta_{\pm}/a_{\pm}$, which is given by
\begin{equation}
\tan\eta_{\pm}=\sinh\left(t\pm \hbar\xi t^2/H\right) ,
\end{equation}
with $\pi/2\geq\eta_+\geq 0$ and $3 \pi/2\geq\eta_-\geq \pi$, the
metrics can be re-expressed in a unitary form as
\begin{equation}
ds_{\pm}^2 = \frac{a_0^2}{\cos^2\eta_{\pm}}\left(-d\eta_{\pm}^2
+\gamma_{\alpha\beta}dx^{\alpha}dx^{\beta}\right) ,
\alpha,\;\beta=1,2,...(d-1) ,
\end{equation}
where $\gamma_{\alpha\beta}$ is the metric for a unit
$(d-1)$-sphere.

We shall consider in what follows the equivalent in our quantum
cosmic scenarios of the static ($d-1$)-dimensional metric. Let us
use the new coordinates
\[x_d =H^{-1}\sin\psi_{d-1}\sin\psi_{d-2}...\sin\psi_2\cos\psi_1\]
\[x_{d-1}
=H^{-1}\sin\psi_{d-1}\sin\psi_{d-2}...\sin\psi_2\sin\psi_1\]
\[x_{d-2} =H^{-1}\sin\psi_{d-1}\sin\psi_{d-2}...\cos\psi_2 \]
\begin{equation}
\end{equation}
\[x_3 =H^{-1}\sin\psi_{d-1}\sin\psi_{d-2}cos\psi_{d-3} \]
\[x_2 =H^{-1}\sin\psi_{d-1}\cos\psi_{d-2} \]
\[x_0=H^{-1}\cos\psi_{d-1}\sinh(H\Theta ') \]
which are defined by $t' \epsilon (-\infty,+\infty)$),
$r\epsilon(0,H^{-1})$, $\psi_{d-1}, \psi_{d-2}, ... \psi_2
\epsilon (0,\pi)$, $\psi_1 \epsilon (0,2\pi)$. These coordinates
will again be referred to either time $\Theta_+ '$ or time
$\Theta_- '$. Setting $r=H^{-1}\sin\psi_{d-1}$, we then find the
metrics
\begin{eqnarray}
&&ds_{\pm}^2 =-\left(1\pm\frac{\hbar\xi t'}{H}\right)^2
dt'^2\left(1-H^2 r^2\right) \nonumber\\ &&+\frac{dr^2}{1-H^2 r^2}
+r^2 d\Omega_{d-2}^2 ,
\end{eqnarray}
where $d\Omega_{d-2}^2$ is the metric on the $(d-2)$-sphere. We
immediately note that this metric is no longer static. The
coordinates defined by that metric cover only the portion of the
spaces with $x_1>0$ and $\sum_{j=2}^d x_j^2 < H^{-2}$ (?), i.e.
the region inside the particle and event horizons of an observer
moving along $r=0$.

Respective instantons can now be obtained by analytically
continuing $\Theta_{\pm}\rightarrow i{\rm T}_{\pm}$ (where we have
taken $\Theta '\equiv \Theta$ for the sake of simplicity in the
expressions), that is $t'\rightarrow i\tau$ and $\xi\rightarrow -
i \chi$, which contain singularities at $r=H^{-1}$, which are only
apparent singularities if ${\rm T}_{\pm}$ are identified with
periods $\pm 2\pi H^{-1}$, or in other words, if $\tau$ is
respectively identified with periods $H(\sqrt{1+8\pi\hbar\chi
H^{-2}}\pm 1)/(2\hbar\chi)$. It follows then that the two spaces
under consideration would respectively behave as though if they
would emit a bath of thermal radiation at the intrinsic
temperatures given by
\begin{equation}
T_{\pm}^{th} =\frac{2\hbar\chi}{H\left(\sqrt{1+8\pi\hbar\chi
H^{-2}}\pm 1\right)} .
\end{equation}
It must be remarked that in the limit when $\chi\rightarrow 0$,
both temperatures $T_{\pm}^{th}$ consistently reduce to the unique
value $H/(2\pi)=\sqrt{\Lambda(d-1)^{-1}}/(2\pi)$, that is the
temperature of a $d$-dimensional de Sitter space [16], even though
$T_-^{th}$ does it more rapidly than $T_+^{th}$ (in fact, for
sufficiently small $\chi$, we can check that $T_-^{th}\simeq
H/(2\pi)$ and $T_+^{th}\simeq\hbar\chi H/(H^2+2\pi\hbar\chi)$).
Note that while we keep $\hbar$ in all definitions concerning the
quantum cosmic spaces, natural units so that $\hbar=G=c=k_B=1$ are
otherwise used when such definitions are used. Now, one can
estimate the entropy of these spaces by taking the inverse to
their temperature. Thus, it can be seen that the entropy of the
universe with scale factor $a_+$ will always be larger than that
for a universe with scale factor $a_-$. It follows then that
whereas the transition from $a_+$ to $a_-$ at $t=0$ would violate
the second law of thermodynamics, the transition from $a_-$ to
$a_+$ at $t=0$ would satisfy it, so making the model with scale
factor $a_+$ evolving along positive time more likely to happen.

The time variables $t$ and $t'$ in Eqns. (3.2), (3.5) and (3.9) do
not admit any bounds other than ($-\infty$, $+\infty$), so that
the involved models can be related with the Barrow's hyper
inflationary model [14], albait the solution $a_+$ here always
respect the second law of thermodynamics because for such a
solution the entropy is an ever increasing function of time [8].

Before closing up this section we shall briefly consider the
static Schwarzschild-quantum mechanically perturbed solutions. It
can be shown that in that case the line element is again not
properly static as they depend on time in their $g_{tt}$
component, that is
\begin{eqnarray}
&&ds_{\pm}^2 =-\left(1\pm\frac{\hbar\xi t'}{H}\right)^2
dt'^2\left(1-\frac{2M}{r}-H^2 r^2\right) \nonumber\\
&&+\frac{dr^2}{1-\frac{2M}{r}-H^2 r^2} +r^2 d\Omega_{d-2}^2 ,
\end{eqnarray}

Instantons for such solutions can also be similarly constructed.
One readily may show that again such instantons describe thermal
baths at given temperatures given now by
\begin{equation}
T_{\pm}^{th} =\frac{2\hbar\chi}{H\left(\sqrt{1+8\pi\hbar\chi
H^{-2}}\pm
1\right)\left(1\mp\frac{2}{3}\epsilon\right)+O\left(\epsilon^2\right)}
.
\end{equation}
where the second sign ambiguity in the denominator refers to the
cosmological (upper) and black hole (lower) horizons and,
according to Ginsparg and Perry [13], $9M^2\Lambda
=1-3\epsilon^2$, with $0\leq \epsilon << 1$, the degenerate case
corresponding just to $\epsilon\rightarrow 0$.

\section{Gravitational waves and semiclassical instability}
In this section we shall restrict ourselves to the solutions
derived in the previous section for just the four-dimensional
case, considering the generation of gravitational waves in the
realm of such solutions and some semiclassical instabilities that
arise when one Euclideanizes ($t\rightarrow i\tau$) the
higher-dimensional solutions. Thus, let us consider first the
tensorial Liftshif-Khalatnikov perturbations corresponding to the
zeroth mode $\ell=0$. From them we can derive [13,16] the
differential equation
\begin{equation}
\nu'' +2\tan\eta\nu' =0 ,
\end{equation}
where $\eta$ and $'=d/d\eta$ refer to the conformal time, either
$\eta_+$ or $\eta_-$, defined in Eq. (3.7). This differential
equation has as general solution
\begin{equation}
\nu= C_0 + C_1 \left(\eta +\frac{1}{2}\sin(2\eta)\right) ,
\end{equation}
where $C_0$ and $C_1$ are given integration constants. We must now
particularize solution (4.2) to be referred to $\eta_{\pm}$. In
the case $\eta_+$ we see that the conformal time runs from $0$
($t=0$) to $\pi/2$ ($t=\infty$). These waves do not destabilize
the space as, though their amplitude does not vanish at the limit
where $\eta_+ \rightarrow\pi/2$, neither it grows with time $t$.
For $\eta_-$ the conformal time runs from $\pi$ ($t=0$ or $t=H^2
/(\hbar\xi)$) to $3\pi/2$ ($t=\infty$). It can be easily seen that
neither these waves can destabilize the space.

For the general case $\ell\neq 0$, we have the general
differential equation, likewise referred to either $\eta_+$ or
$\eta_-$,
\begin{equation}
\nu'' +2\tan\eta\nu'+\ell(\ell+2)\nu =0 .
\end{equation}
The solution to this differential equation can be expressed as
\begin{equation}
\nu =\cos^3\eta C_{\ell-1}^{(2)}(\sin\eta) ,
\end{equation}
with $C_{n}^{(\alpha)}$ the ultraspherical (Gegenbauer)
polynomials of degree 2. Now, for $\eta_+ =0$ or $\eta_- =\pi$,
the amplitude vanishes for even $\ell = 2, 4, 6, ...$, and becomes
\[\nu = (-1)^{(\ell-1)/2}\frac{\Gamma
\left(2+\frac{\ell-1}{2}\right)}{\Gamma(2)\left(\frac{\ell-1}{2}\right)!}
,\] for odd $\ell=1, 3, 5, ...$. For $\eta_+ =\pi/2$, $\nu=
(\ell+2)!/[6(l-1)!]$ and for $\eta_- =3\pi/2$,
$\nu=(-1)^{\ell-1}(\ell+2)!/[6(l-1)!]$. Once again the considered
spaces are therefore stable to tensorial perturbations for nonzero
$\ell$. It is worth mentioning that for the solution corresponding
to $\eta_-$ and even $\ell$, the absolute value of the amplitude
of the gravitational waves would first increase from zero (at
$t=0$) to reach a maximum value at $t=H/(2\hbar\xi)$, to then
decrease down to zero at $t=H/(\hbar\xi)$, and finally steadily
increase all the time to reach its final finite value of unit
order as $t\rightarrow\infty$. Clearly. a distinctive
observational effect predicted by that cosmic model would be the
generation of gravitational waves whose amplitude adjusted to the
given pattern.

A general derivation of Eqns. (4.1) and (4.3) from a general
traceless rank-two tensor harmonics which is an eigenfunction of
the Laplace operator on $S^3$ and satisfies the eigenvalue
equation $\nabla_a\nabla^a H_{cd}^{(n)}=-(n^2 -3)H_{cd}^{(n)}$ can
be found in Refs. [13,16].

We add finally some comments to the possibility that our closed
spaces develop a semiclassical instability. We shall use the
Euclidean approach. In order to see if our Euclideanized solutions
are stable or correspond to semiclassical instabilities, it will
suffice to determine the eigenvalues of the differential operator
[13,21]
\begin{equation}
G_{abcd}\Phi^{ab}\simeq -\Box\Phi_{cd}-2R_{acbd}\Phi^{ab}\simeq
\lambda\Phi_{cd} ,
\end{equation}
where $\Phi_{ab}$ is a metric perturbation. Now, if all
$\lambda\geq 0$, the Euclideanized spaces are stable, showing a
semiclassical instability otherwise. Stability can most readily be
shown if, by analytically continuing metrics (3.10), the metric on
the ($d-2$)-sphere, $d\Omega_{d-2}$, turns out to be expressible
as the Kahler metric associated to a 2-sphere. Thus, let us
introduce the complex transformation
\begin{equation}
Z=2\tan\left(\psi_{d-2}/2\right)\exp\left(i\int
d\Omega_{d-3}\right) ,
\end{equation}
and hence in fact we can derive
\begin{equation}
d\Omega_{d-2}=\frac{d\bar{Z}dZ}{\left(1+\frac{1}{4}\bar{Z}Z\right)^2}
\end{equation}
and the Kahler potential
\begin{equation}
K=2\log\left(1+\frac{1}{4}\bar{Z}Z\right) ,
\end{equation}
so showing that, quite similarly to what it happens in the
$d$-dimensional de Sitter space, the instantons constructed from
metrics (3.10) are stable. Whether or not a space-time
corresponding to Schwarzschild-generalized de Sitter metric would
show a semiclassical instability is a question that required
further developments and calculations.

\section{conclusions and comments}
This paper deals with new four-dimensional and $d$-dimensional
cosmological models describing an accelerating universe in the
spatially flat and closed cases. The ingredients used for
constructing these solutions are minimal as they only specify a
cosmic relativistic field described by just Hilbert-Einstein
gravity and the notion of the quantum entanglement of the
universe, that is the probabilistic quantum effects associated
with the general matter content existing in the universe or its
generalization for the closed cases. Two of such models correspond
to an equation of state $p=w\rho$ with parameter $w<-1$ for their
entire evolution, and still other of them which covers a period in
its future also with $w<-1$; that is to say, these three solutions
are associated with the so-called phantom sector, showing however
a future evolution of the universe which is free from most of the
problems confronted by usual phantom scenarios; namely, violent
instabilities, future singularities, incompatibility with the
previous existence of a matter-dominated phase, classical
violations of energy conditions or inadequacy of the holographic
description. Therefore we also denote such quantum cosmic models
as {\it benigner phantom} models. All these models can be regarded
as generalizations or perturbations of the either exponential or
hyperbolic form of the de Sitter space. The hyperbolic solution
are given in a $d$-dimensional manifold which is particularized in
the four-dimensional case in the Euclideanized extension that
allowed us to derive quantum formulas for the temperature that
reduce to that of Gibbons-Hawking when the perturbation is made to
vanish. Finally, the generation of gravitational waves in some of
the considered models has been studied in the realm of the
Lifshitz-Khalatnikov perturbation formalism for the spatially
closed case. It is also shown that none of these waves destabilize
the space-time, as neither the vector and scalar cosmological
perturbations do in the spatially flat and closed cases.

\acknowledgments

\noindent This work was supported by MEC under Research Project
FIS2008-06332/FIS. The authors benefited from discussions with
C.L. Sig\"{u}enza.

\end{document}